\documentclass[manuscript]{aastex}
\usepackage{graphicx}
\usepackage{epstopdf}
\usepackage{natbib}
\usepackage{amssymb}
\usepackage{amsmath}

\begin{document}
\title{Six-year Optical Monitoring of BL Lacertae Object \\1ES 0806+52.4}
\author{Zhongyi Man, Xiaoyuan Zhang, and Jianghua Wu}
\affil{Department of Astronomy, Beijing Normal University,
Beijing 100875, China;}
\email{jhwu@bnu.edu.cn}
\author{Xu Zhou}
\affil{Key Laboratory of Optical Astronomy, National Astronomical Observatories,
Chinese Academy of Sciences, 20A Datun Road, Beijing 100012, China}
\author{Qirong Yuan}
\affil{Department of Physics and Institute of Theoretical Physics, Nanjing Normal University,
Nanjing 210046, China}

\begin{abstract}
We present the results of the first systematic long-term multi-color optical monitoring of the BL Lacertae object 1ES 0806+52.4. The monitoring was performed in multiple passbands with a 60/90 cm Schmidt telescope from December 2005 to February 2011. The overall brightness of this object decreased from 2005 December to 2008 December, and regained after that. A sharp outburst probably occurred around the end of our monitoring program. Overlapped on the long-term trend are some short-term small-amplitude oscillations. No intra-night variability was found in the object, which is in accord with the historical observations before 2005. By investigating the color behavior, we found strong bluer-when-brighter chromatism for the long-term variability of 1ES 0806+52.4. The total amplitudes at the {\it c, i,} and {\it o} bands are 1.18, 1.12, and 1.02 mags, respectively. The amplitudes tend to increase toward shorter wavelength, which may be the major cause of bluer-when-brighter. Such bluer-when-brighter is also found in other blazars like S5 0716+714, OJ 287, etc. The hard X-ray data collected from the {\it Swift}/BAT archive was correlated with our optical data. No positive result was found, the reason of which may be that the hard X-ray flux is a combination of the synchrotron and inverse Compton emission but with different timescales and cadences under the leptonic Synchrotron-Self-Compton (SSC) model.
\end{abstract}

\section{Introduction}

Active galactic
nuclei (AGN) is powered by a supermassive black hole surrounded by an accretion
disk. Blazar, one of the most violently variable class of AGN, is distinguished by large and rapid flux variability \citep{2012ApJ...756...13B,2014A&A...562A..79S}, high and variable polarization
\citep{2008ApJ...672...40H, 2008PASJ...60L..37S, 2014ApJ...781L...4G}, and basically a
non-thermal continuum. Depending on the presence of strong emission lines in spectrum, a blazar can be further termed either as a flat-spectrum radio quasar (FSRQ) or a BL Lacerate object.

Two typical components can be seen in the spectral energy
distribution (SED) of blazars: One at low frequencies from radio to UV
or soft X-rays, the other ranges from X-rays to gamma-rays (e.g. \citealt{2001AIPC..558..346T,2012ApJ...756...13B}). The low energy component is believed to be caused by the synchrotron process of the
relativistic particles in the jet. The high energy one, however, is interpreted as either due to inverse
Compton scattering of the low energy photons by the same ensemble of relativistic
electrons responsible for the synchrotron emission named Synchrotron–Self-Compton (SSC) model (e.g. \citealt{2003APh....18..593M,2002MNRAS.336..721K}), or due to the hadronic processes initiated by synchrotron radiation of protons co-accelerated with the electrons in the jet (hardronic models e.g. \citealt{2001APh....15..121M, 2006IJMPA..21.6015L}).

1ES 0806+52.4 was identified as a BL Lac object \citep{1993ApJ...412..541S},
based on the observations carried out on both radio band from
the Green Bank 91m telescope \citep{1991ApJS...75....1B} and X-ray bands from the Einstein Slew Survey
\citep{1992ApJS...80..257E}. The redshift of its host galaxy  is estimated as z = 0.138
\citep{1998A&A...334..459B}. Very high energy gamma-ray observations of 1ES
0806+52.4 was reported by the Whipple Collaboration \citep{2003ApJ...599..909D,2004ApJ...603...51H}, the HEGRA Collaboration \citep{2004A&A...421..529A} and VERITAS \citep{2002APh....17..221W}. {\it R}-band optical observations of this BL Lac object were conducted from 2000 to 2001 \citep{2002xsac.conf..301K}, and from 2002 to 2004 \citep{2009MNRAS.398..832K}. The former highlighted the variability with one magnitude on long time scales \citep{2002xsac.conf..301K}. Till now, no long-term systematic optical monitoring has been conducted on this object, and no research has been made on the correlation between radiation at different wavelengths.

Our quasi-simultaneous  multi-color monitoring of 1ES 0806+52.4 was carried out from 2005 December to 2011 February.
In order to investigate the correlation between the optical and the hard X-ray emission, X-ray data covering the same period were collected from the {\it Swift}/BAT data archive and were analyzed with the optical data.

\section{Observation and Data Reduction}

Our observations were performed on a 60/90 cm Schmidt telescope at the Xinglong Station of the National Astronomical Observatories of China (NAOC). Mounted at the telescope main focus was a Ford Aerospace 2048 $\times$ 2048 thick CCD camera which was then surrogated by a new E2V 4096$ \times$ 4096 CCD in 2006 December. The field of view is enlarged from 58\arcmin $\times$ 58 \arcmin  to 96\arcmin $\times$ 96\arcmin. The resolution is increased from 1.\arcsec7 pixel$^{-1}$ to 1.\arcsec3 pixel$^{-1}$. The blue quantum efficiency of the new E2V thinned CCD is also higher than the former thick one. The telescope is equipped with a 15 color intermediate-band filters, covering a wavelength range from 3000 to 10,000 $\textrm{\AA}$ \citep{1996AJ....112..628F}. Four F sub-dwarfs \citep{1983ApJ...266..713O} are used as the standard stars to calibrate the magnitude. For the details of the filter system, definition of the corresponding magnitudes and how to derive flux from the magnitude, one can refer to \citet{2000PASP..112..691Y, 2001ChJAA...1..372Z,2003A&A...397..361Z}.

Our monitoring program on 1ES 0806+52.4 started on the night of 2005 December $12^{th}$, and ended on 2011 February $11^{th}$. After ruling out the nights with bad weather condition and those for other targets, the actual number of observation nights is 77. Filters in {\it e, i,} and {\it m} bands were adopted in 2005 - 2006. During most of the nights in this period, more than ten exposures, sometimes up to twenty, were made in each band with a temporal resolution about 20 minutes. Filters were then surrogated by {\it c, i,} and {\it o} bands from the end of 2006 December to 2011 February, and less than ten exposures were made on most nights. The exposure time determined by the moon phase, weather condition and filter varies from 60 to 540 seconds. The central wavelengths of {\it c, e, i, m,} and {\it o} bands are 4210, 4920, 6660, 8020, and 9190 \AA, respectively. The {\it i} magnitude can be transformed into the broadband R magnitude for stars by using the formula given by \citet{2003A&A...397..361Z} and for AGNs by using the formula given by \citet{2013ApJS..204...22D}.

We did bias subtraction and flat-fielding using the BATC automatic data reduction software PIPELINE I \citep {1996AJ....112..628F, 1999AJ....117.2757Z}. Four stars, C1, C2, C3, and C4 (see Fig. 1), were selected as reference stars \citep {1998PASP..110..105F}, and we chose star C5 as check star (for a reasonable selection of reference and check stars, see \citealt{1988AJ.....95..247H}). We extracted the instrumental aperture magnitudes of the blazar and the five comparison stars from each frame by using the algorithm of \citet {1987PASP...99..191S}. The radii of the aperture and the sky annuli were adopted as 3, 7, and 10 pixels, respectively, and they may be adjusted slightly according to the seeing. The brightness of the blazar was measured relative to the average magnitude of the four reference stars. The accuracy of our measurement is shown by the check star's differential magnitude, the difference between the magnitude of C5 and the average magnitude of the four reference stars. We observed stars C1, C2, C3, C4, C5 on a photometric night, and used the photometric F-type standard stars mentioned above to calibrate their standard magnitudes. The results are listed in Table~\ref{tb1}.

\section{Light Curve}

The samples of observational log and results are given in Table~\ref{tb2} -~\ref{tb6}. The columns are observation date and time in universal time, exposure time in seconds, Julian date, magnitude, error, and differential magnitude of star C5 (its nightly averages were set to zero).

We gathered 1399 optical data points in total: numbers of photometric points for {\it c, e, i, m,} and {\it o} bands are 292, 150, 490, 171 and 296, respectively. The light curves of the overall monitoring period in five bands are shown in Figure 2. There are basically two outbursts with a concave shape between them. The peak of the first outburst occurred on 2006 Mar $15^{th}$ (JD 2,453,810), and then the brightness of the object started falling until 2008 Dec $11^{th}$ (JD 2,454,812). Later on, the curves turned up to the second outburst which is more powerful than the first one. No details can be known about the strong outburst because our observation ceased on 2011 Feb $11^{th}$ (JD 2,455,604). Overlapped on the long-term large-amplitude variations are some short-term small-amplitude oscillations during JD 2,454,000 to JD 2,455,000. We also notice that a double-peak structure of 1ES 0806+52.4 was spotted during 2002-2004 \citep{2009MNRAS.398..832K}, yet it was not found in our observations.

The overall amplitudes in {\it e, i}, and {\it m} bands from 2005 December $15^{th}$ to 2006 November $19^{th}$ are 0.17, 0.12, and 0.09 mags, respectively, the amplitudes in {\it c, i,} and {\it o} band from 2006 December $4^{th}$ to 2011 February $11^{th}$ are 1.18, 1.12, and 1.02 mags. We note that the overall amplitudes tend to decline with the decreasing frequency. We examined the intra-night variability of this object by using the methods suggested by \citet{2010AJ....139.1269D}, but eventually found no obvious intra-night variability during the whole time span. It is in accord with the observation from 2002 to 2004 \citep{2009MNRAS.398..832K}, which detected no intra-night variability for this source either. Nevertheless, our average observational duration on each night is only 3.6 hours, which may not be long enough to detect the intra-night variability, if there are some cases. Some blazars display strong intra-night variability. For example, S5 0716+714 was observed to vary by 0.117 mags in 1.1 hrs \citep{2013ApJS..204...22D}. PKS 2155-304 was detected a very fast variability rate of 0.43 mag/h in the observation session \citep{2014A&A...562A..79S}.

\section{Color Behavior}

The long-term color behavior of 1ES 0806+52.4 was investigated based on our data. We adopted the {\em e} and {\em m} bands to
calculate the color in the first stage from 2005 to 2006, and used the {\em c} and {\em o} bands in the second stage from 2006 to 2011. It is because the differences in wavelength of the two pairs of filters are relatively large, and can thereby denote a better spectral shape. The color-magnitude diagrams are displayed in Figure 3.

The linear least square method with measurement errors taken into consideration
was used to fit the points. The points in the left panel for the first stage have a linear regression of $y=0.537x-7.847$, and the correlation coefficient is 0.808, which means a significance level of 95\%. The points in the right panel for the second stage have a linear regression of $y=0.176x-1.791$, and the correlation coefficient is 0.792, which also means a significance level of 95\%. Both
panels show strong bluer-when-brighter (BWB) chromatism. The intra-night color behavior of 1ES 0806+52.4 was not investigated due to the absence of intra-night variability.

A number of BL Lac objects have
been reported displaying the BWB phenomenon so far. Significant BWB correlations were found for both inter-night and intra-night variations in the typical BL Lac object S5 0714+716
(e.g.,\citealt{2005AJ....129.1818W}). Inter-night BWB was found in BL Lacerate (e.g.,\citealt{2012A&A...538A.125Z}). In OJ 287, a well studied BL Lac, BWB is also found on long time scale \citep{2011AJ....141...65D}. In fact, such BWB chromatism is likely to
be a general feature of BL Lac objects (e.g. \citealt{2003ApJ...590..123V,2010MNRAS.404.1992R}). For FSRQ, there were once claims of redder-when-brighter trend (e.g. \citealt{2006A&A...450...39G,2006MNRAS.371.1243H}). However, \citet{2011A&A...534A..59G} found this color behavior for only one object in a sample of 29 FSRQs. The mechanism of BWB was simulated \citep{2011AJ....141...65D}, and the result shows that BWB is probably due to the difference of the amplitude at different wavelengths. The variation amplitude of 1ES 0806+52.4 was found to decrease with decreasing frequency, which can at least partly explain the BWB behaviour observed in this object.

\section{Cross-correlation Analysis and Discussions}

In order to research the emission at different wavelengths, our optical data was compared with the time series data provided by {\it Swift}/Burst Alert Telescope (BAT), which has
monitored 1ES 0806+52.4 in hard X-ray band since 2005. The {\it Swift}/BAT hard X-ray transient monitor provides near real-time coverage
of the X-ray sky in the energy range 15-50 keV, or
$3.63\times10^{18}-1.21\times10^{19}$ Hz in frequency.
The BAT observes 88\% of the sky each day with a detection sensitivity of 5.3 mCrab
 for a full-day observation and a time resolution as fine as 64 seconds
\citep{2013ApJS..209...14K}. We used the daily average data offered by {\it Swift}/BAT. Both sequences were transformed into flux so that we can analyze them together. The light curves of i band and X-ray band are shown in Figure 4.

We used the z-transformed discrete correlation functions (ZDCFs;
\citealt{1997ASSL..218..163A}) to search for the optical-hard X-ray correlations. The ZDCF
differs from the discrete correlation function (DCF; \citealt{1988ApJ...333..646E})
for it bins the data points into equal population bins and uses
Fisher’s z-transform to stabilize the highly skewed distribution of the
correlation coefficient. It is much more efficient than the DCF in uncovering
correlations involving the variability timescale, and deals with under-sampled
light curves better than both the DCF and the interpolated cross-correlation
function (ICCF; \citealt{1987ApJS...65....1G}). The results are displayed in Figure 5 for all the data from JD 2,453,717 to JD 2,455,604.

In Figure 5, no obvious peak of the DCF values can be found, indicating an
absence of correlation between the two bands. We try to explain the result as follows:

Firstly, there are daily data for X-ray band, yet the optical observation has not been taken regularly: 77 daily average points in 1187 days in total. The time resolution of our optical data is not compatible with that of {\it Swift}/BAT, which may possibly result in the loss of important fluctuations in the optical light curve, especially at the end of our monitoring program. It is probably hard to find a strong correlation between two samples with very different sampling rates. Secondly, as shown in Figure 4, no significant flare or outburst can be found in both optical and X-ray bands (There may be a strong flare at the end of the optical data but the measurements are too few. The X-ray data obviously have no corresponding flare). As a result, it might be hard to find any correlation.

Thirdly, provided that the no-correlation result is not caused by the limited data sampling in optical band or the lack of significant variability event in both bands, then it could be a reflect of a complex radiation mechanism. We notice that the hard X-ray band from $3.63\times10^{18}$ Hz to $1.21\times10^{19}$ Hz lies in the transition area of the two peaks in the SED of 1ES 0806+52.4 \citep{2008AIPC.1085..403C}. Emission in this regime may be a combination of the synchrotron and inverse Compton mechanisms. Variations from the synchrotron and inverse Compton processes may have different amplitudes and cadences, hence a combination of them may show complicated variability and can hardly be found correlated with the optical variation. A similar explanation was raised to frame why the correlation between X-ray and other bands cannot be found in blazar S5 0716+714 \citep{2013A&A...552A..11R}.

Besides, there are also some other blazars showing no correlation between X-ray and other bands, which might be caused by the different location of emission regions in different bands. For instance, 3C 279 was observed by \citet{2010Natur.463..919A}, and its X-ray emission has not been found accompanied quasi-simultaneously by the optical/GeV flares. An argument is that X-ray photons are produced further down to the jet compared to optical – GeV photons. The case that the variations in X-ray band is not correlated with that of optical/gamma-ray (so-called orphan gamma-ray flare) band was also found in other blazars such as Mrk 501 \citep{2012A&A...541A..31N}, 1ES 1959+650 \citep{2005ApJ...621..181D}, etc. The TeV flare in Mrk 501 may be produced by an electromagnetic cascade initiated by very-high-energy gamma-rays in the intergalactic medium \citep{2012A&A...541A..31N}, and a Hadronic Synchrotron Mirror Model was employed to explain the orphan TeV flare of 1ES 1959+650 \citep{2005ApJ...621..181D}. These results, together with ours, can be used to investigate the detail of the variability of blazars, i.e. the radiation mechanism, the location of the emission region, etc.

\section{Conclusions}

Our monitoring work targeting at the BL Lac object 1ES 0806+52.4 was
 conducted in five intermediate optical wavebands from 2005 December to
2011 February using the 60/90 cm Schmidt telescope located at the Xinglong Station of the National Astronomical Observatories of China (NAOC). It was the first systematic multi-color optical monitoring of 1ES 0806+52.4 on long timescale, and could thereby be effectively used to further study both long and short term flux and spectral variability of this object.

The analysis of the data reveals its overall brightness declined from 2005 December to 2008 December, and it turned brighter during 2009 to 2011. Overlapped on the long-term trend are some short-term small-amplitude oscillations. Unfortunately, our data samples are not abundant considering the long period of time we covered, and there are several interruptions in our observation. No obvious intra-night variability was found during the monitoring period, and the historical observation from 2002-2004 revealed the same feature. We did not find the double-peak structure as the former observation has suggested. Color behavior on long timescale was studied, and strong BWB phenomenon is revealed. This is consistent with the color behavior of most blazars. No obvious correlation was found between our optical variations and hard X-ray variability collected from the {\it Swift}/BAT archive. This result can be attributed to the different time resolution of the optical and X-ray bands or lack of significant variability event in the observation session; moreover, it may suggest the hard X-ray emission is a combination of both synchrotron radiation and inverse Compton scattering.

We gathered 1399 data points in total for 1ES 0806+52.4, which is the largest optical multi-color database for the variability of the object todate. Our data can also be correlated with the data other than the hard X-ray wavelength, e.g., radio, Gamma-ray, etc. for a more comprehensive investigation of the broad-band behavior of 1ES 0806+52.4.

We thank the anonymous referee for insightful comments and constructive
suggestions, and we also thank Yang Chen for his advice and help to us. Our work has been supported by National Basic Research Program of China 973 Program 2013CB834900, Chinese National Natural Science Foundation grant 11273006, 11173016, 11073023 and the Scientific Research Foundation of Beijing Normal University.

\bibliography{mybib}
\bibliographystyle{apj}

\begin{deluxetable}{rrrrrr}
\tablecolumns{6}
\tablewidth{0pc}
\tablecaption{Magnitude of 5 Comparison Stars.\label{tb1}}
\tablehead{
\colhead{band} & \colhead{C1} & \colhead{C2} & \colhead{C3} & \colhead{C4} &
\colhead{C5}
}
\startdata
\emph{c} & 14.035 & 15.410 & 15.379 & 16.498 & 16.187 \\
\emph{e} & 13.291 & 14.859 & 14.938 & 15.901 & 15.809 \\
\emph{i} & 12.653 & 14.349 & 14.534 & 15.412 & 15.548 \\
\emph{m} & 12.476 & 14.216 & 14.420 & 15.245 & 15.256 \\
\emph{o} & 12.387 & 14.164 & 14.402 & 15.170 & 15.399 \\
\enddata
\end{deluxetable}

\begin{deluxetable}{rrrrrrr}
\tabletypesize{\footnotesize}
\tablecolumns{7}
\tablewidth{0pc}
\tablecaption{Observational Log and Results in the $c$ Band\label{tb2}}
\tablehead{
\colhead{Observation Date} & \colhead{Observation Time} & \colhead{Exposure Time}
& \colhead{Julian Date} & \colhead{$c$} & \colhead{\emph{$c_{err}$}} & \colhead{dfmagC5}\\
\colhead{(UT)} & \colhead{(UT)} & \colhead{(s)} & \colhead{JD} & \colhead{(mag)}
& \colhead{(mag)} & \colhead{(mag)}
    }
\startdata
 2006 12 04 & 18:44:51.0 & 480 & 2454074.28115 & 15.886 & 0.034 & 0.064\\
 2006 12 04 & 19:07:27.0 & 480 & 2454074.29684 & 15.806 & 0.035 & -0.011\\
 2006 12 04 & 19:29:52.0 & 480 & 2454074.31241 & 15.809 & 0.031 & 0.008\\
 2006 12 04 & 19:52:18.0 & 480 & 2454074.32799 & 15.832 & 0.032 & 0.012\\
 2006 12 04 & 20:14:53.0 & 480 & 2454074.34367 & 15.850 & 0.037 & -0.006\\
\enddata
\tablecomments{This table is available in its entirety in machine-readable and Virtual Observatory (VO) forms in the online journal. A portion is shown here for guidance
regarding its form and content.}
\end{deluxetable}

\begin{deluxetable}{rrrrrrr}
\tabletypesize{\footnotesize}
\tablecolumns{7}
\tablewidth{0pc}
\tablecaption{Observational Log and Results in the $e$ Band\label{tb3}}
\tablehead{
\colhead{Observation Date} & \colhead{Observation Time} & \colhead{Exposure Time}
& \colhead{Julian Date} & \colhead{$e$} & \colhead{\emph{$e_{err}$}} & \colhead{dfmagC5}\\
\colhead{(UT)} & \colhead{(UT)} & \colhead{(s)} & \colhead{JD} & \colhead{(mag)}
& \colhead{(mag)} & \colhead{(mag)}
}
\startdata
 2005 12 15 & 15:01:46.0 & 300 & 2453720.12623 & 15.991 & 0.085 & -0.024\\
 2005 12 15 & 15:17:43.0 & 300 & 2453720.13730 & 15.993 & 0.078 & 0.106\\
 2005 12 15 & 15:34:01.0 & 300 & 2453720.14862 & 15.925 & 0.081 & -0.056\\
 2005 12 15 & 15:49:56.0 & 300 & 2453720.15968 & 15.894 & 0.083 & -0.053\\
 2005 12 15 & 16:05:54.0 & 300 & 2453720.17076 & 15.865 & 0.071 & -0.071\\
\enddata
\tablecomments{This table is available in its entirety in machine-readable and Virtual Observatory (VO) forms in the online journal. A portion is shown here for guidance
regarding its form and content.}
\end{deluxetable}

\begin{deluxetable}{rrrrrrr}
\tabletypesize{\footnotesize}
\tablecolumns{7}
\tablewidth{0pc}
\tablecaption{Observational Log and Results in the $i$ Band\label{tb4}}
\tablehead{
\colhead{Observation Date} & \colhead{Observation Time} & \colhead{Exposure Time}
& \colhead{Julian Date} & \colhead{$i$} & \colhead{\emph{$i_{err}$}} & \colhead{dfmagC5}\\
\colhead{(UT)} & \colhead{(UT)} & \colhead{(s)} & \colhead{JD} & \colhead{(mag)}
& \colhead{(mag)} & \colhead{(mag)}
}
\startdata
 2005 12 12 & 15:44:16.0 & 180 & 2453717.15574 & 15.488 & 0.034 & -0.021\\
 2005 12 12 & 15:55:15.0 & 180 & 2453717.16337 & 15.514 & 0.033 &  0.006\\
 2005 12 12 & 16:05:08.0 & 180 & 2453717.17023 & 15.408 & 0.030 & -0.004\\
 2005 12 12 & 16:15:18.0 & 180 & 2453717.17729 & 15.416 & 0.030 & -0.024\\
 2005 12 12 & 16:25:24.0 & 180 & 2453717.18431 & 15.408 & 0.028 &  0.042\\
\enddata
\tablecomments{This table is available in its entirety in machine-readable and Virtual Observatory (VO) forms in the online journal. A portion is shown here for guidance
regarding its form and content.}
\end{deluxetable}

\begin{deluxetable}{rrrrrrr}
\tabletypesize{\footnotesize}
\tablecolumns{7}
\tablewidth{0pc}
\tablecaption{Observational Log and Results in the $m$ Band}\label{tb5}
\tablehead{
\colhead{Observation Date} & \colhead{Observation Time} & \colhead{Exposure Time}
& \colhead{Julian Date} & \colhead{$m$} & \colhead{\emph{$m_{err}$}} & \colhead{dfmagC5}\\
\colhead{(UT)} & \colhead{(UT)} & \colhead{(s)} & \colhead{JD} & \colhead{(mag)}
& \colhead{(mag)} & \colhead{(mag)}
}
\startdata
 2005 12 12 & 15:49:15.0 & 300 & 2453717.15920 & 15.248 & 0.044 &  0.014\\
 2005 12 12 & 16:00:10.0 & 300 & 2453717.16678 & 15.238 & 0.041 & -0.057\\
 2005 12 12 & 16:10:21.0 & 300 & 2453717.17385 & 15.189 & 0.040 &  0.053\\
 2005 12 12 & 16:20:26.0 & 300 & 2453717.18086 & 15.236 & 0.045 & -0.019\\
 2005 12 12 & 16:30:24.0 & 300 & 2453717.18778 & 15.264 & 0.051 & -0.006\\
\enddata
\tablecomments{This table is available in its entirety in machine-readable and Virtual Observatory (VO) forms in the online journal. A portion is shown here for guidance
regarding its form and content.}
\end{deluxetable}

\begin{deluxetable}{rrrrrrr}
\tabletypesize{\footnotesize}
\tablecolumns{7}
\tablewidth{0pc}
\tablecaption{Observational Log and Results in the $o$ Band\label{tb6}}
\tablehead{
\colhead{Observation Date} & \colhead{Observation Time} & \colhead{Exposure Time}
& \colhead{Julian Date} & \colhead{$o$} & \colhead{\emph{$o_{err}$}} & \colhead{dfmagC5}\\
\colhead{(UT)} & \colhead{(UT)} & \colhead{(s)} & \colhead{JD} & \colhead{(mag)}
& \colhead{(mag)} & \colhead{(mag)}
}
\startdata
 2006 12 04 & 18:57:10.0 & 480 & 2454074.28970 & 14.818 & 0.027 & -0.063\\
 2006 12 04 & 19:20:16.0 & 540 & 2454074.30574 & 14.840 & 0.028 & -0.040\\
 2006 12 04 & 19:42:41.0 & 540 & 2454074.32131 & 14.822 & 0.026 & -0.026\\
 2006 12 04 & 20:05:09.0 & 540 & 2454074.33691 & 14.809 & 0.026 & -0.009\\
 2006 12 04 & 20:27:41.0 & 540 & 2454074.35256 & 14.840 & 0.027 &  0.006\\
\enddata
\tablecomments{This table is available in its entirety in machine-readable and Virtual Observatory (VO) forms in the online journal. A portion is shown here for guidance
regarding its form and content.}
\end{deluxetable}

\begin{figure}
\plotone{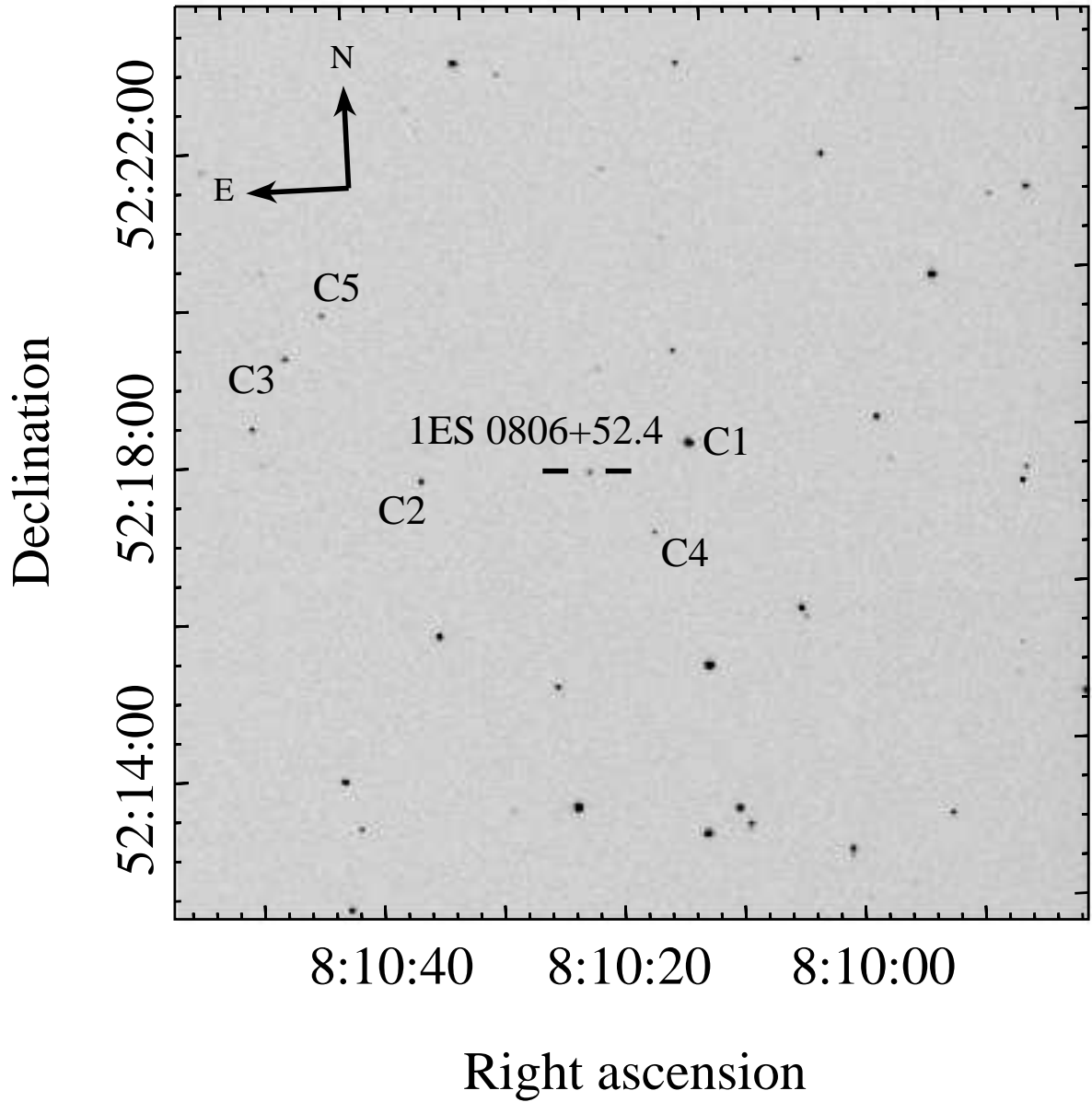}
\caption{Finding chart of 1ES 0806+52.4 and the 5 comparison stars taken in \emph{i} band with the
60/90 Schmidt telescope on Feb $5^{th}$, 2007. The size is 12\arcmin $\times$ 12\arcmin  (or 512 $\times$ 512 in pixels). }
\end{figure}

\begin{figure}
\plotone{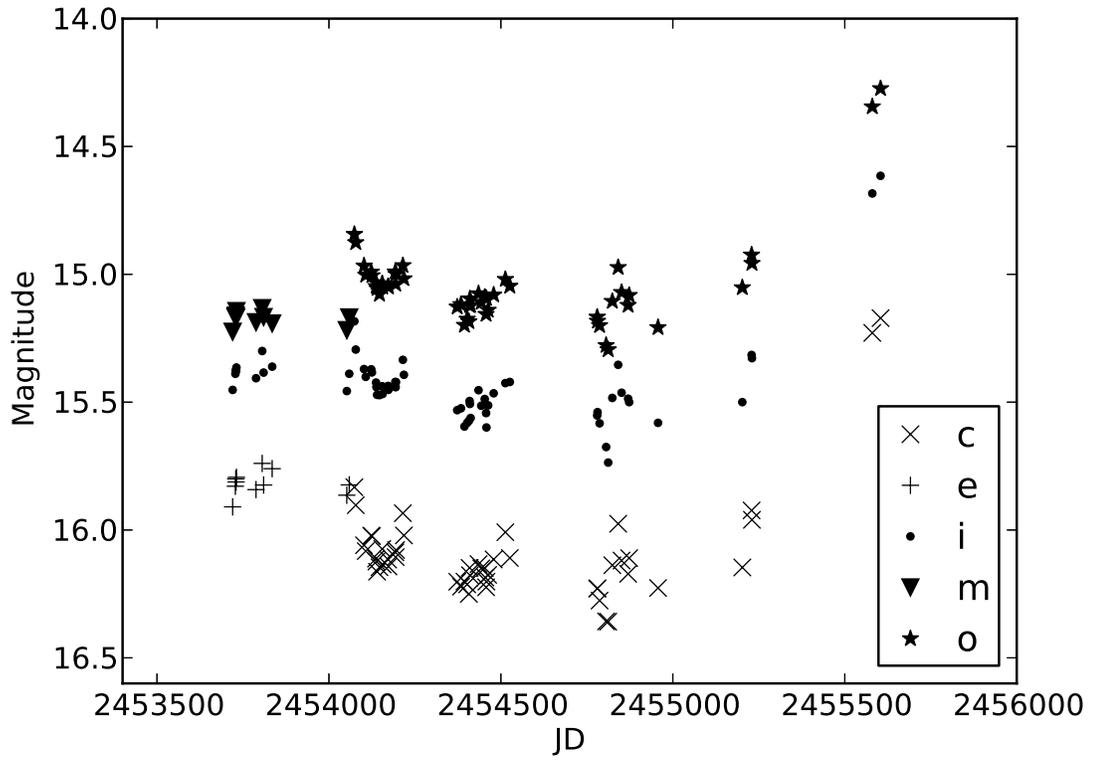}
\caption{Nightly average light curves of 1ES 0806+52.4 in the \emph{c}, \emph{e}, \emph{i},
\emph{m}, and \emph{o} bands.}
\end{figure}

\begin{figure}
\epsscale{0.8}
\plotone{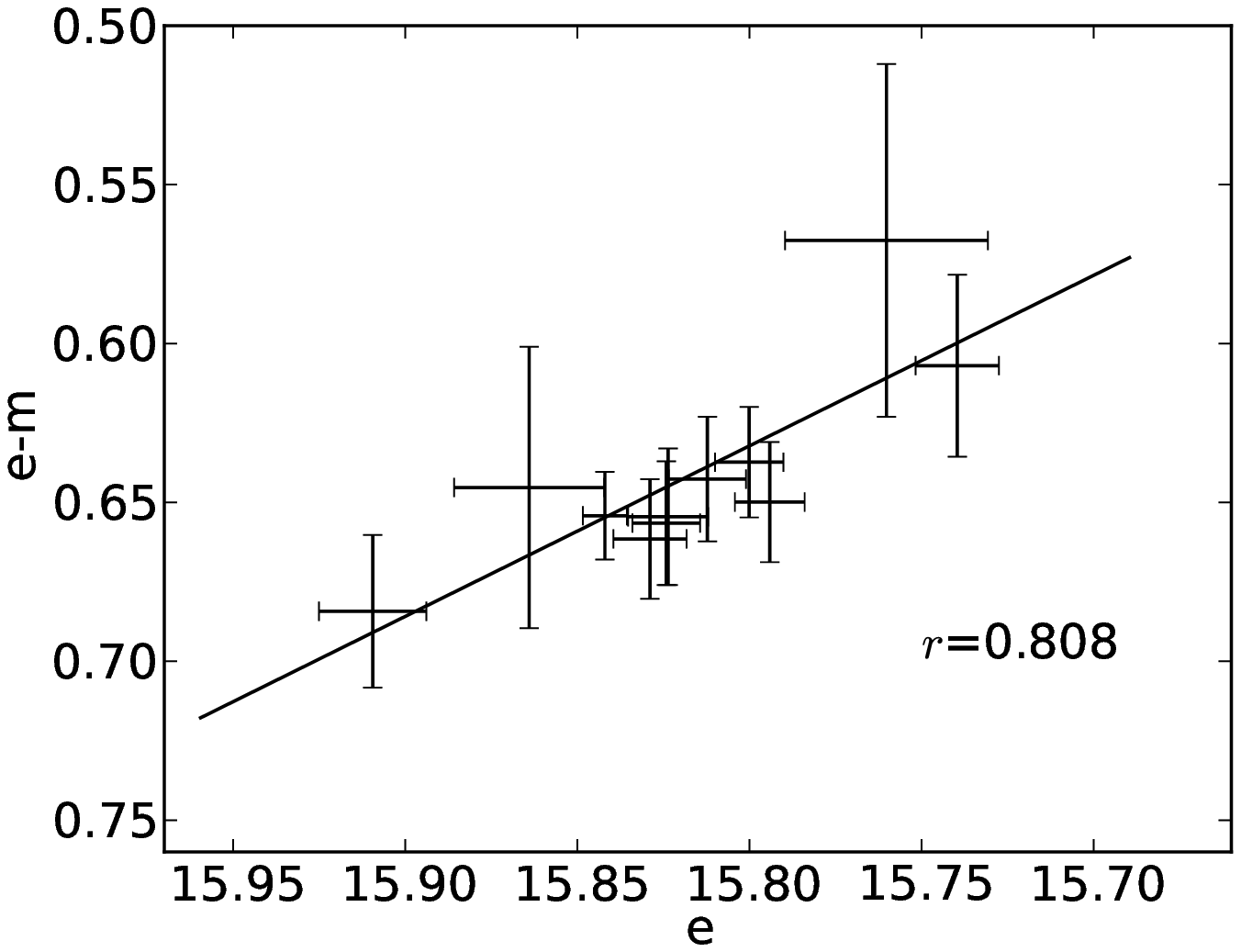}
\plotone{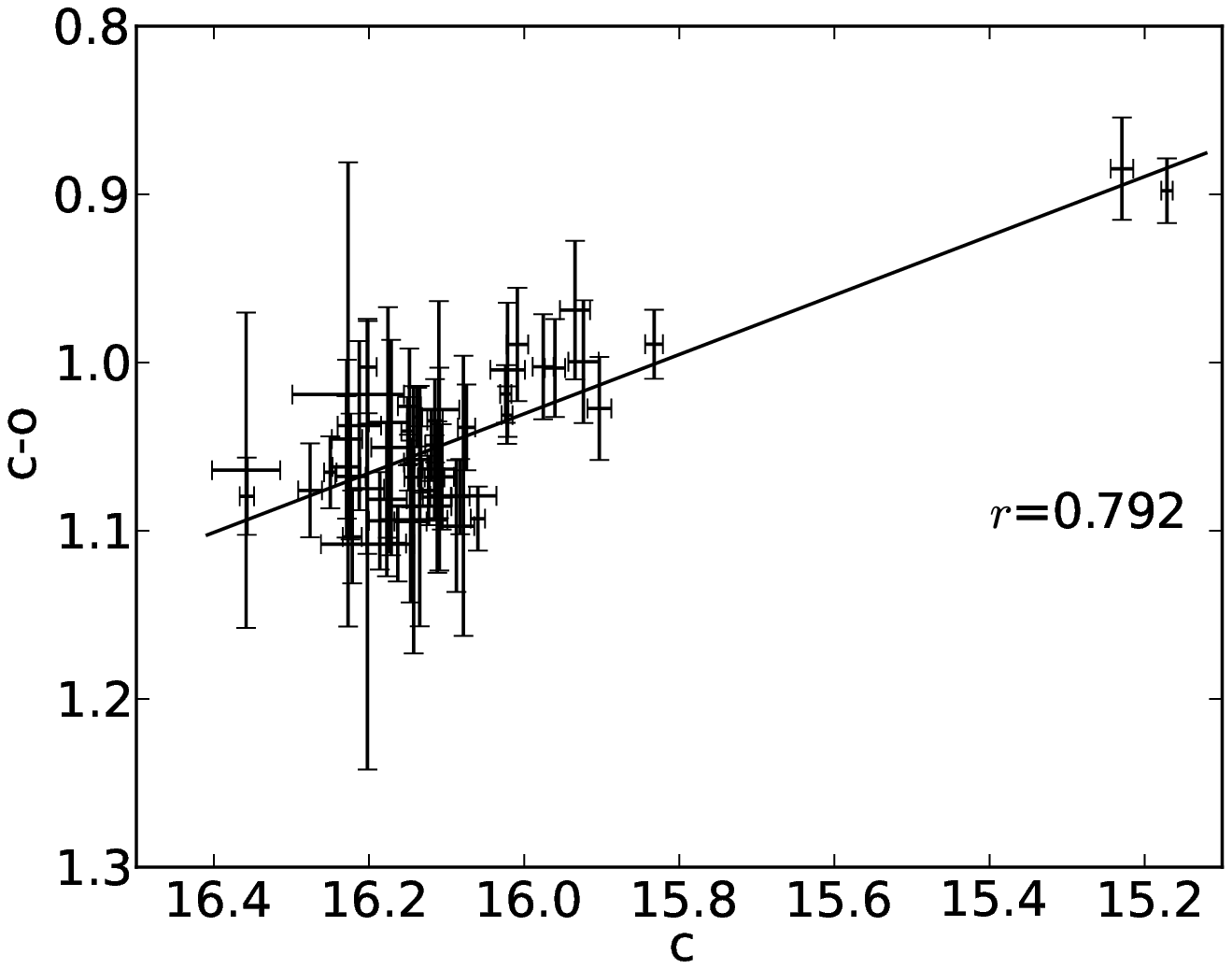}
\caption{Long term color behavior of the first (up) and the second (bottom) stages. The nightly-average data are used. Strong bluer-when-brighter chromatisms can be found.  }
\end{figure}

\begin{figure}
\plotone{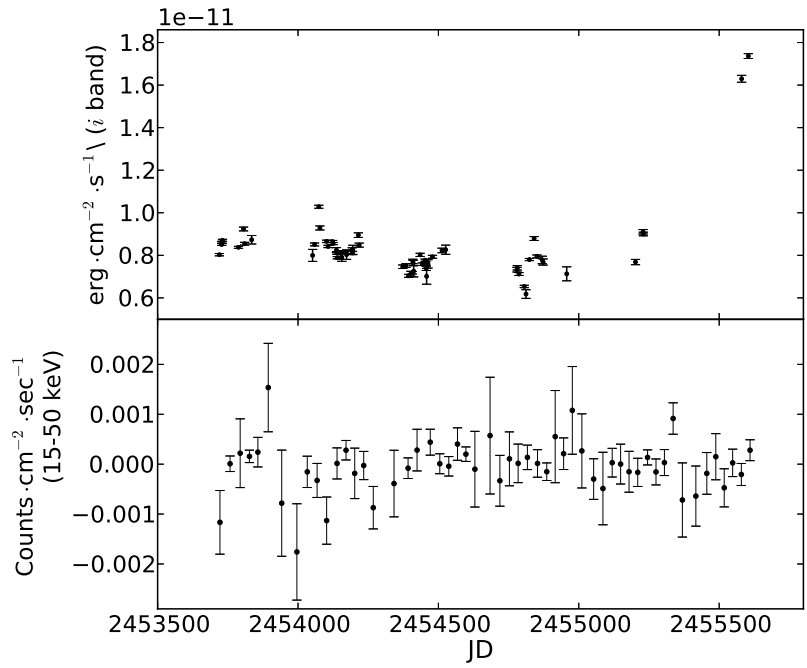}
\caption{Light curve in \emph{i} band and X-ray band are shown here. Monthly average X-ray data were presented instead of daily average to provide a clearer view of the general variation in the graph.}
\end{figure}

\begin{figure}
\plotone{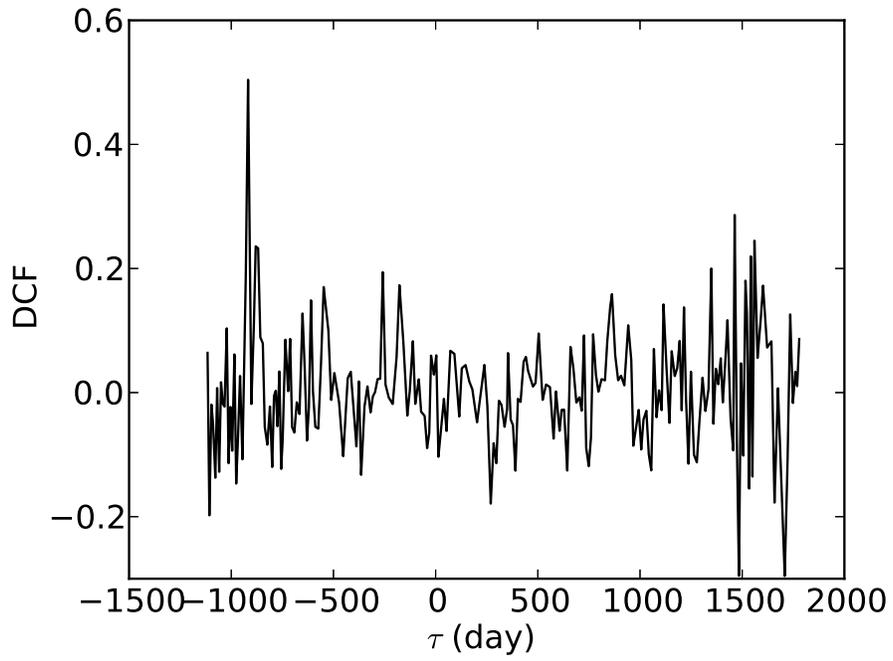}
\caption{The DCF of the z-transformed discrete correlation function.}
\end{figure}

\end{document}